\documentclass[aps,prx,twocolumn,nopacs,superscriptaddress]{revtex4-1}

\usepackage[breaklinks=true,colorlinks,citecolor=blue,linkcolor=blue,urlcolor=blue]{hyperref}
\usepackage{epsfig,mathrsfs,color,latexsym,subfigure,marginnote,graphicx,verbatim,relsize,mathrsfs,color,array,amsmath,amsfonts,amssymb,graphicx}
\usepackage{xspace}
\newcommand{\SB}[1] {{\it{\color{blue}#1}}}

\newcommand{\tss}[1]{\ensuremath{^{\text{#1}}}}
\newcommand{\rrscan}{r\tss{2}SCAN\xspace}
\newcommand{\veck}{\boldsymbol{k}}
\newcommand\vecq{\boldsymbol{q}}
\newcommand{\eph}{\textit{e-ph}}

\begin{document}

\title{Magnetism-Enhanced Strong Electron-Phonon Coupling in Infinite-Layer Nickelates}

\author{Ruiqi~Zhang}
\thanks{These authors contributed equally to this work}
\email[Corresponding author:~]{rzhang16@tulane.edu}
\affiliation{Department of Physics and Engineering Physics, Tulane University, New Orleans, LA 70118, USA}

\author{Yanyong~Wang}
\thanks{These authors contributed equally to this work}
\affiliation{Department of Physics and Engineering Physics, Tulane University, New Orleans, LA 70118, USA}

\author{Manuel~Engel}
\thanks{These authors contributed equally to this work}
\affiliation{VASP Software GmbH, Berggasse 21/14, 1090 Vienna, Austria}

\author{Christopher Lane}
\affiliation{Theoretical Division, Los Alamos National Laboratory, Los Alamos, New Mexico 87545, USA}

\author{Henrique Miranda}
\affiliation{VASP Software GmbH, Berggasse 21/14, 1090 Vienna, Austria}

\author{Lin Hou}
\affiliation{Department of Physics and Engineering Physics, Tulane University, New Orleans, LA 70118, USA}
\affiliation{Theoretical Division, Los Alamos National Laboratory, Los Alamos, New Mexico 87545, USA}

\author{Sugata Chowdhury}
\affiliation{Department of Physics and Astronomy, Howard University, Washington DC 20059, USA}

\author{Bahadur Singh}
\affiliation{Department of Condensed Matter Physics and Materials Science, Tata Institute of Fundamental Research, Mumbai 400005, India}

\author{Bernardo~Barbiellini}
\affiliation{Department of Physics, School of Engineering Science, LUT University, FI-53850 Lappeenranta, Finland}
\affiliation{Department of Physics, Northeastern University, Boston, MA 02115, USA}
\affiliation{Quantum Materials and Sensing Institute, Northeastern University, Burlington, MA 01803, USA}

\author{Jian-Xin Zhu}
\affiliation{Theoretical Division, Los Alamos National Laboratory, Los Alamos, New Mexico 87545, USA}

\author{Robert S. Markiewicz}
\affiliation{Department of Physics, Northeastern University, Boston, MA 02115, USA}
\affiliation{Quantum Materials and Sensing Institute, Northeastern University, Burlington, MA 01803, USA}

\author{E. K. U. Gross}
\affiliation{Fritz Haber Center for Molecular Dynamics, Institute of Chemistry, The Hebrew University of Jerusalem, Jerusalem 91904, Israel}

\author{Georg Kresse}
\affiliation{VASP Software GmbH, Berggasse 21/14, 1090 Vienna, Austria}
\affiliation{University of Vienna, Faculty of Physics, Kolingasse 14-16, A-1090 Vienna, Austria}

\author{Arun~Bansil}
\email[Corresponding author:~]{ar.bansil@neu.edu}
\affiliation{Department of Physics, Northeastern University, Boston, MA 02115, USA}
\affiliation{Quantum Materials and Sensing Institute, Northeastern University, Burlington, MA 01803, USA}

\author{Jianwei Sun}
\email[Corresponding author:~]{jsun@tulane.edu}
\affiliation{Department of Physics and Engineering Physics, Tulane University, New Orleans, LA 70118, USA}

\begin{abstract}

Intriguing analogies between the nickelates and the cuprates provide a promising avenue for unraveling the microscopic mechanisms underlying high-$T_c$ superconductivity. While electron correlation effects in the nickelates have been extensively studied, the role of electron-phonon coupling (EPC) remains highly controversial. Here, by taking pristine LaNiO$_2$ as an exemplar nickelate, we present an in-depth study of EPC for both the non-magnetic (NM) and the $C$-type antiferromagnetic ($C$-AFM) phase using advanced density functional theory methods without invoking $U$ or other free parameters. The weak EPC strength $\lambda$ in the NM phase is found to be greatly enhanced ($\sim$4$\times$) due to the presence of magnetism in the $C$-AFM phase. This enhancement arises from strong interactions between the flat bands associated with the Ni-3$d_{z^2}$ orbitals and the low-frequency phonon modes driven by the vibrations of Ni and La atoms. The resulting phonon softening is shown to yield a distinctive kink in the electronic structure around 15 meV, which would provide an experimentally testable signature of our predictions.  Our study highlights the critical role of local magnetic moments and interply EPC in the nickelate. 
 \end{abstract} 

\maketitle

\SB{Introduction}\textbf{\---}High-$T_c$ superconductors, such as cuprates, exhibit rich and complex phase diagrams, attracting significant interest for exploring new physics paradigms and advancing next-generation quantum technologies~\cite{Tranquada1996,Kievelson2003,Bertinshaw2019}. Despite over four decades of extensive studies, the pairing mechanism underlying high-$T_c$ superconductivity remains elusive. Comparative studies between cuprates and analogous materials can yield valuable insights into their fascinating properties.  In this context, the recent discovery of superconductivity in the family of hole-doped infinite-layer (IL) nickelates, RNiO$_2$ (R = La, Pr, Nd), with a 3$d^9$ electronic filling similar to Cu$^{2+}$ in the cuprates, has ignited intense interest in these materials~\cite{Hepting2020,Goodge2021,Zhang2021_nickelate,ZhangPRLLaNiO2,Wu2020,Wang2020,Lu2021science,Rossi_nat_phy,OsadaAM2021,Pickett2021,Tam2021,Fowlie2022,Lee2023_linear_tc,SC_dome_NNO,Phase_D_NdNiO2,Chen_nickelate_xray,Norman2020,Botana2020,Karp2020,BeenPRX2021,Sawatzky2019,Li2019a}. Additionally, superconductivity has been observed under moderate pressure or strain in bi-layer and tri-layer nickelates structurally similar to hole-doped cuprates, further highlighting the relevance of nickelates to high-$T_c$ research~\cite{SunLa3Ni2O72023,Zhu2024Tlayer,Wang2024BLPRLNO}. Consequently, exploring the similarities and differences between nickelates and cuprates has become a central focus in this field. 

The IL-nickelates, in particular, exhibit numerous unconventional phenomena analogous to cuprates, such as strange metallicity of the normal state near optimal doping~\cite{Lee2023_linear_tc}, a similar superconducting dome~\cite{SC_dome_NNO,Phase_D_NdNiO2}, and strong magnetic exchange interactions~\cite{Lu2021science}, spin-glass behaviors~\cite{Lin_2022,OrtizPRR2022} among other features. Despite these signatures of unconventional pairing mechanisms, low-energy kinks observed in angle-resolved photoemission spectroscopy (ARPES) experiments~\cite{Sun_ARPES_LNO} indicative of strong electron-phonon coupling (EPC) effects in the IL-nickelates. Additionally, a recent experimental has linked reduced superexchange interactions under compressive strain to potentially strong EPC in IL-nickelates~\cite{Gao_starin_magnetic_2024}. Thus, a comprehensive exploration of EPC's role in these materials is essential.

A density functional theory (DFT) based calculation~\cite{Nomura_eph_LNO} reported a very small EPC strength $\lambda$ of $\sim$0.2 in the non-magnetic (NM) phase of IL-nickelates, suggesting that the role of EPC in the nickelates could be safely ignored. While a larger enhanced EPC strength, by a factor of 2-3, was found by including GW perturbation theory corrections~\cite{Two_gap_Louie}, a later, more accurate GW calculation including dynamical correlations~\cite{GW_weak_LNO}, showed that the EPC enhancement is quite small and remains far too weak to account for the experimentally observed $T_c$ values. However, the NM phase is found to be much higher in energy than a number of low-energy phases with local spin moments, including several antiferromagnetic and stripe phases~\cite{Zhang2021_nickelate,ZhangPRLLaNiO2}. Note that many experimental studies have indicated the absence of long-range order in the parent LaNiO$_2$~\cite{Hayward1999,Hayward2003,Lu2021science,OrtizPRR2022,ZhaoPRL2021}, but with magnetism showing  strong AFM interactions~\cite{Li2019a, Fowlie2022, Lu2021science,Gao_starin_magnetic_2024, 2021arXiv211113668O}, indicating the presence of competing low-energy phases those can frustrate long--range order, leaving behind a pseudogap similar to the case of the doped cuprates~\cite{ZhangYPANS}. Additionally, magnetic phases show a fully gapped Ni-$d_{x^2-y^2}$ band, consistent with the observed Mott-Hubbard bands~\cite{Tam2021,Goodge2021,Chen_nickelate_xray}, whereas no splitting of the Ni-$d_{x^2-y^2}$ band is predicted in the NM phase calculations~\cite{Zhang2021_nickelate}.  A large EPC strength $\lambda$ was predicted in the magnetic phase of the nickelates based on frozen-phonon calculations, but only bond disproportionating modes were considered~\cite{CO_elph_Julien,COBO_elph_Julien,carrascoalvarez:tel-04504289}. Therefore, a systematic investigation of EPC is critically needed, where the spin, orbital, charge, and lattice degrees of freedom should be treated on an equal footing across the entire Brillouin zone (BZ).

\begin{figure*}[t!]
\centering
\includegraphics[width=0.95\textwidth]{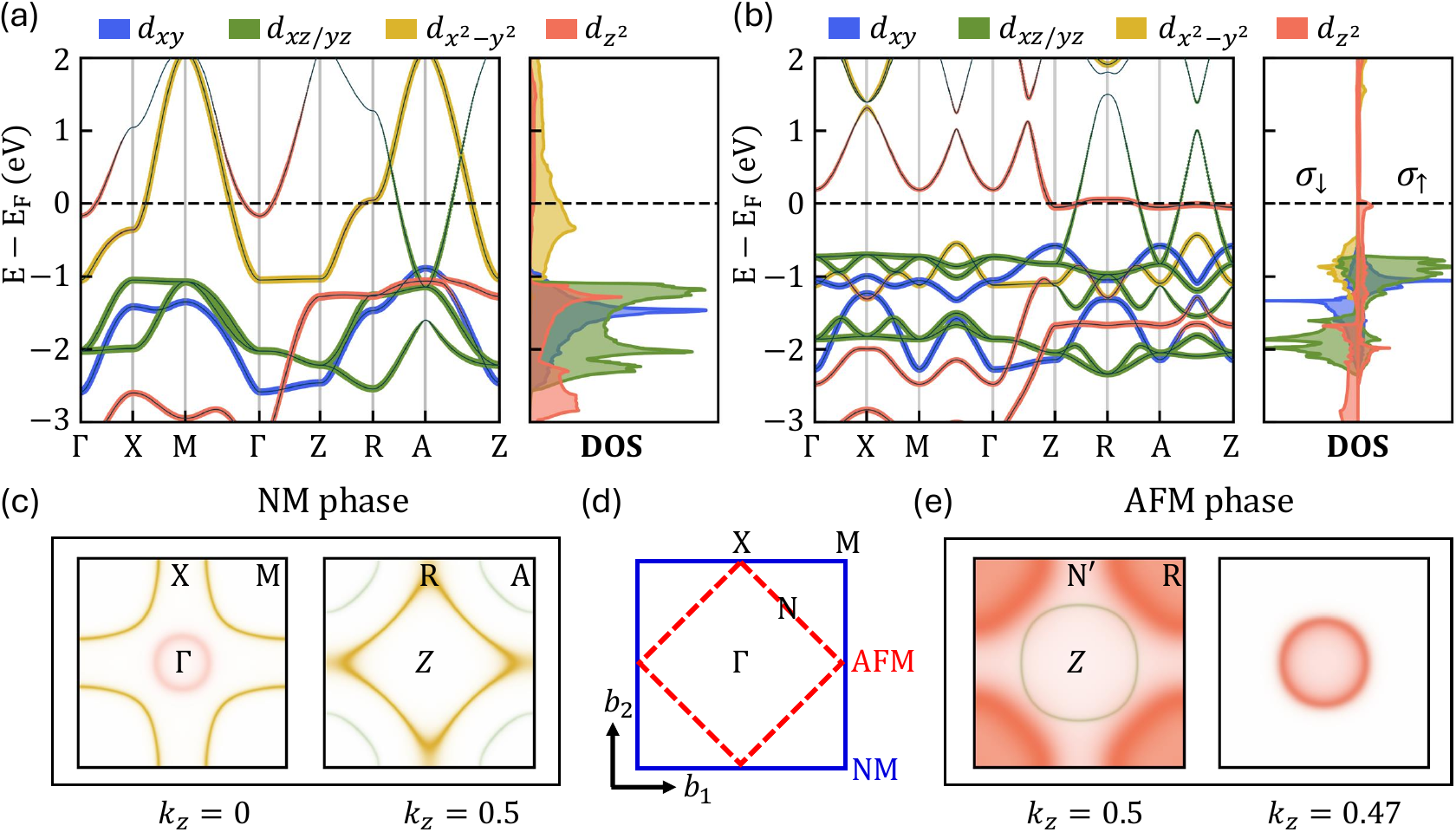}
\caption{(a) Calculated band structure of the NM phase of LaNiO$_2$ along the high-symmetry lines in the first Brillouin Zone (BZ). Contributions of various Ni $d$-orbitals to the bands are identified. High-symmetry points $\Gamma$, X, and M lie in the $k_z =0$ plane, while Z, A, and R are the corresponding points in the $k_z = \pi/c$ plane. Side panel shows the partial density of states (PDOS) from the Ni 3$d$ orbitals. (b) Same as (a), but for the $C$-AFM phase, using the same $k$-path as in the NM case. Side panel shows the site-projected, orbitally resolved PDOS for Ni sites with positive magnetic moments. Positive (negative) values represent spin-up (spin-down) contributions to the DOS. (c) Two-dimensional plot of the FS of the NM phase on the $k_z=0$ plane (right panel) and the $k_z = \pi/c$ plane (left panel). (d) Schematic of the NM and AFM BZs in the $k_z = 0$ plane, with high-symmetry points marked. (e) Two-dimensional plot of the FS of the $C$-AFM phase in the $k_z = \pi/c$ plane (left panel) and the $k_z = 0.47$ plane (right panel) in the $C$-AFM BZ.}
\label{fig:elec}
\end{figure*}

In this work, we present a comprehensive investigation of EPC in both the NM and $C$-type antiferromagnetic ($C$-AFM) phases of pristine LaNiO$_2$~\cite{Zhang2021_nickelate,ZhangPRLLaNiO2}, by using the inputs from advanced \rrscan~functional~\cite{R2SCANJPCL}. Our previous study has shown that several stripe phases are slightly lower in energy than the $C$ -AFM phase~\cite{ZhangPRLLaNiO2}. However, performing EPC calculations in these stripe phases with lower symmetry is impractical because of the highly demanding computational cost for a large supercell. However, given the similar electronic structures between stripe and $C$-AFM phases~\cite{ZhangPRLLaNiO2}, we consider the $C$-AFM phase to be a suitable representative for EPC studies in LaNiO$_2$. Our results demonstrate significant enhancement of EPC in the $C$-AFM phase due to the presence of low-energy flat bands predominantly arising from the Ni-3$d{z^2}$ orbital~\cite{Zhang2021_nickelate,ZhangPRLLaNiO2,Choi_flat_band}. Furthermore, phonon dispersions in the $C$-AFM phase exhibit significant softening relative to the NM phase,  driven by pronounced Fermi Surface (FS) nesting and EPC effects. The predicted EPC strength $\lambda$ for $C$-AFM phase is approximately 0.66, which is four times greater than the NM phase value of 0.16. This notable enhancement arises from strong coupling between flat bands and low-frequency phonon modes involving La and Ni vibrations, rather than from high-frequency breathing modes of oxygen atoms. Beyond strong electron correlations, our findings highlight the importance of EPC in the nickelates ~\cite{ZhangPRLLaNiO2} and call for further experimental and theoretical studies in this direction.

\SB{Results and Discussion}

\SB{Electronic structure of NM and $C$-AFM phases}\textbf{\---}The parent NM LaNiO$_{2}$ shares the crystal structure with the parent infinite-layer CaCuO$_2$ [space group $P4/mmm$ ($\#123, D_{4h}$)]. The unit cell of the $C$-AFM phase is constructed using a $\sqrt{2} \times \sqrt{2} \times 1$ supercell of NM LaNiO$_{2}$, where the magnetic interactions of the nearest-neighbor Ni atoms are antiferromagnetic within the $xy$-plane and ferromagnetic along the $z$-direction. Previous studies have shown that the \rrscan functional gives similar results to the SCAN functional but with greater numerical stability~\cite{R2SCANJPCL}. We therefore adopted the \rrscan functional in our calculations. Additional computational details can be found in the Methods section. The calculated magnetic moment of Ni atoms from \rrscan is approximately $0.9\mu_{B}$, slightly smaller than the SCAN result of $1.0\mu_{B}$~\cite{Zhang2021_nickelate}. Indeed, the predicted magnetic moments of Ni from \rrscan are in good agreement with previous DFT+$U$~\cite{Choi_flat_band} and DMFT calculations~\cite{Leonov2020}.  Consistent with earlier DFT calculations~\cite{Botana2020,Zhang2021_nickelate}, the band structure of the NM phase computed using \rrscan is shown in Fig.~\ref{fig:elec}(a), where two bands are seen crossing the Fermi level. One band is primarily contributed by the half-filled Ni 3$d_{x^2-y^2}$ orbital, while the other arises from the strong hybridization between Ni (3$d_{z^2}$, 3$d_{xz/yz}$) and La $5d$ orbitals. The right and left panels of Fig.~\ref{fig:elec}(c) show the projections of the FS of the NM phase onto the $k_z = 0$ and $k_{z}=\pi/c$ planes, respectively. In Fig.~\ref{fig:elec}(c), an electron pocket primarily contributed by the Ni 3$d_{z^2}$ orbital is observed around the $\Gamma$ point. And, the half-filled Ni 3$d_{x^2-y^2}$ orbital forms an open hole-pocket centered at the M point on the $k_z = 0$ plane and a closed electron-pocket centered at the Z point on the $k_z = \pi/c$ plane. Additionally, the Ni 3$d_{xz/yz}$ orbital creates another electron pocket centered at the A point on the $k_z = \pi/c$ plane. These observations are consistent with previous studies. Further discussion of the FS characteristics can be found in our earlier publication~\cite{Zhang2021_nickelate}.

\begin{figure*}[t!]
\centering
\includegraphics[width=0.95\textwidth]{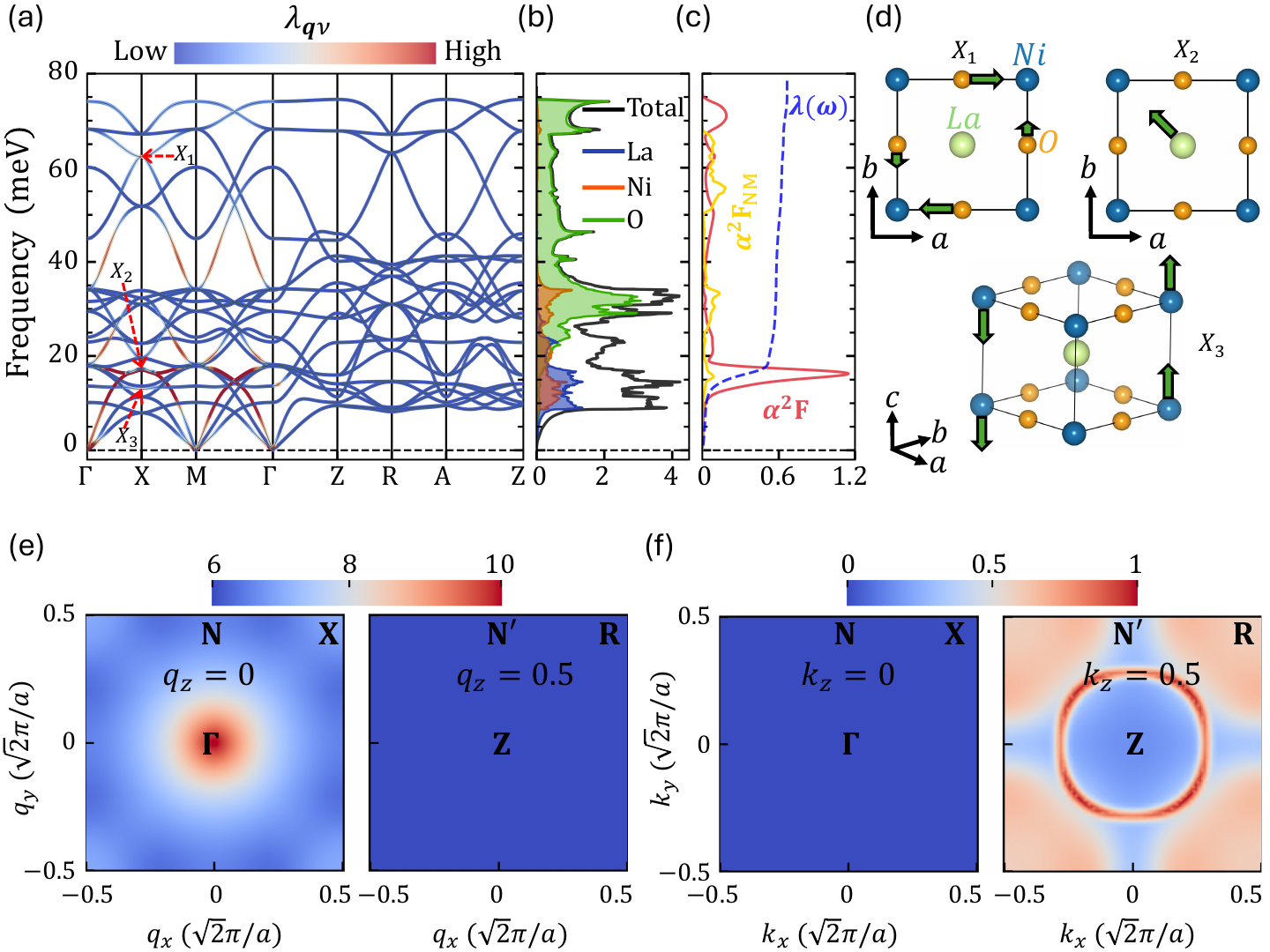}
\caption{(a) Unfolded phonon dispersion of the $C$-AFM phase along the high-symmetry lines in the first NM BZ. Marker color intensity indicates the magnitude of the EPC strength, $\lambda_{\textbf{q}\nu}$, for the $C$-AFM phase. (b) Atom-projected phonon density of states (PDOS). (c) Calculated Eliashberg spectral functions, $\alpha^2F(\omega)$ (solid red lines), and cumulative EPC strength, $\lambda$ (dashed blue lines), for the $C$-AFM phase. The NM $\alpha^2F$ (solid orange line) is included for comparison.  A $k$-mesh of $60 \times 60 \times 60$ and a $q$-mesh of $30 \times 30 \times 30$ were used for full BZ sampling. The $\alpha^2F(\omega)$ of NM phase is replotted here for comparison. (d) Side view of selected phonon modes at high-symmetry points, with green arrows representing the direction of atomic vibrations. (e) Calculated Fermi nesting function for $q_z = 0$ (left panel) and $q_z = \pi/c$ (right panel). (f) Distribution of $k$-resolved EPC strength, $\lambda_{\veck}$, defined as $\lambda_{\veck} = \sum_{mn,\nu\vecq} |g_{mn,\nu}({\veck}, \vecq)|^2 \delta(\varepsilon_{m\textbf{k}+\textbf{q}} - \varepsilon_\text{F}) \delta(\varepsilon_{n\textbf{k}} - \varepsilon_\text{F}) \times (2 / \omega_{\textbf{q}\nu})$, across the $k_x - k_y$ plane at $k_z = 0$ (left panel) and $k_z = \pi/c$ (right panel).}
\label{fig:afm_epc}
\end{figure*}

Figure~\ref{fig:elec}(b) shows the folded band structure of the $C$-AFM phase within the NM BZ. A significant splitting of the Ni 3$d_{x^2-y^2}$ band of approximately 2 eV is observed, which is consistent with our earlier SCAN results~\cite{Zhang2021_nickelate}, previous DFT+$U$ calculations~\cite{Choi_flat_band}, and the Hubbard band-splitting reported in experiments~\cite{Tam2021,Goodge2021,Chen_nickelate_xray}. More importantly, our \rrscan calculations exhibit remarkably flat bands composed of Ni 3$d_{z^2}$ orbitals pinned at the Fermi level on the $k_{z}=\pi/c$ plane, resulting in a high-order van Hove singularity (VHS) in the density of states (DOS), see the right panel of Fig.~\ref{fig:elec}(b). These results align well with previous SCAN~\cite{Zhang2021_nickelate,ZhangPRLLaNiO2} and DFT+$U$ calculations~\cite{Choi_flat_band,Kuwei_NNO}. The flat bands generate quasi-2D FS near the $k_{z}=\pi/c$ plane [Fig.~\ref{fig:afm_epc}(e)]. Selected projections of the FS on the $k_z=0.5$ and $k_z=0.47$ planes are displayed in the left and right panels of Fig.~\ref{fig:afm_epc}(e), respectively. A hole pocket contributed by the Ni 3$d_{xz/yz}$ orbital is observed at the Z point on the $k_z=0.5$ plane. Additionally, a broadened FS, primarily from the Ni 3$d_{z^2}$ orbital, appears around the R point on the $k_z=0.5$ plane, indicating its flat nature. On the $k_z=0.47$ plane, only the flat FS from the Ni 3$d_{z^2}$ orbital is observed, suggesting that the FS in the $C$-AFM phase is nearly confined around the $k_z=0.5$ plane. These results suggest that the low-energy physics of the infinite-layer nickelates is characterized by multi-orbitals, as noted in previous studies~\cite{Zhang2021_nickelate,ZhangPRLLaNiO2,Choi_flat_band}.

\begin{figure*}[t!]
\centering
\includegraphics[width=0.95\textwidth]{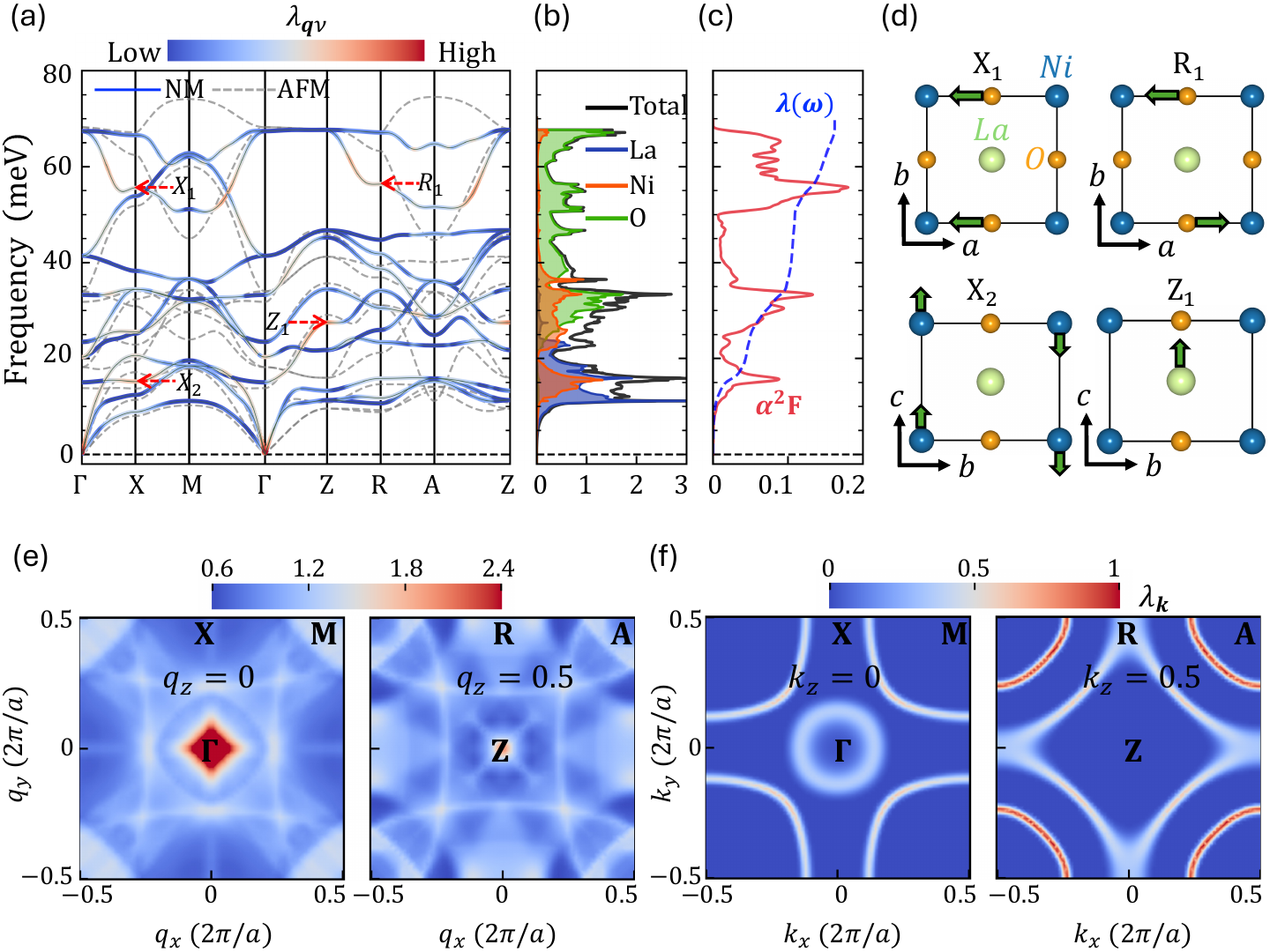}
\caption{(a) Calculated phonon dispersion in the NM phase (solid blue lines) along high-symmetry lines in the first BZ. Marker color-intensity indicates the value of the EPC strength, $\lambda_{\textbf{q}\nu}$, for the NM phase. For comparison, the unfolded phonon dispersions of the $C$-AFM phase (dashed gray lines) are also shown. (b) Atom-projected phonon density of states (PDOS). (c) Calculated Eliashberg spectral functions, $\alpha^2F(\omega)$ (solid red lines), and the cumulative EPC strength, $\lambda$ (dashed blue lines), for the NM phase. A $k$-mesh of $60 \times 60 \times 60$ and a $q$-mesh of $30 \times 30 \times 30$ were used for full BZ sampling. (d) Side view of selected phonon modes at high-symmetry points, with green arrows representing the direction of atomic vibrations. Blue, green, and orange balls represent La, Ni, and O atoms, respectively.
(e) Calculated Fermi nesting function for $q_z = 0$ (left panel) and $q_z = \pi/c$ (right panel). (f) Distribution of $k$-resolved EPC strength, $\lambda_{\veck}$, defined as $\lambda_{\veck} = \sum_{mn,\nu\vecq} |g_{mn,\nu}({\veck}, \vecq)|^2 \delta(\varepsilon_{m\textbf{k}+\textbf{q}} - \varepsilon_\text{F}) \delta(\varepsilon_{n\textbf{k}} - \varepsilon_\text{F}) \times (2 / \omega_{\textbf{q}\nu})$, across the $k_x - k_y$ plane at $k_z = 0$ (left panel) and $k_z = \pi/c$ (right panel).}
\label{fig:nm_epc}
\end{figure*}

\SB{Evaluation of EPC for the NM phase}\textbf{\---}Having discussed the electronic structures of the IL nickelates, we turn to consider phonon dispersions and EPC for the NM and $C$-AFM phases with reference to Fig.~\ref{fig:nm_epc}(a). A key difference between the two phases is the pronounced phonon softening observed in $C$-AFM relative to the NM phase, although the highest oxygen full-breathing mode becomes hardened in $C$-AFM~\cite{Hardening_YWang}. Specifically, the quadrupolar mode of oxygen atoms, with a frequency of $51.1\ \text{meV}$ in the NM phase, softens to $45.0\ \text{meV}$ in the $C$-AFM phase. This substantial reduction indicates the influence of magnetic ordering on lattice vibrations and enhanced \eph~interactions. In contrast, the full-breathing mode, which has a frequency of $62.8\ \text{meV}$ in the NM phase, hardens to $74.1\ \text{meV}$ in the $C$-AFM phase. This hardening suggests that magnetic interactions suppress certain phonon-mediated electron interactions, thereby stiffening this high-frequency mode. These results highlight that local magnetic moments are crucial for understanding the lattice vibrations of LaNiO$_2$ and must be included in EPC studies. The pronounced phonon softening observed in the $C$-AFM phase is primarily driven by two key factors: (i) Significant FS nesting from $k_z = 0$ plane phonons, and (ii) enhanced EPC due to the presence of flat bands near the Fermi level in the $C$-AFM phase. We defer these discussions to the following sections.

To identify which phonon modes are critical for EPC in the NM phase, we plot the phonon dispersion with phonon mode-resolved EPC strengths $\lambda{\textbf{q}\nu}$ along high-symmetry lines [Fig.~\ref{fig:nm_epc}(a)]. From the corresponding total phonon density of states (PhDOS) and partial PhDOS, decomposed by atomic contributions [Fig.~\ref{fig:nm_epc}(b)], we can see that low-frequency modes below 20 meV are predominantly contributed by La and Ni atoms. It can be observed that vibrations of strongly coupled La and Ni atoms form flat bands in the 10–15 meV range, accompanied by van Hove singularities (VHSs) in the PhDOS. In the intermediate frequency range of 20–40 meV, phonon vibrations are primarily contributed by Ni and O atoms, whereas high-frequency modes above 40 meV are dominated by the lighter O atoms.  We also identify two phonon flat bands above 40 meV. The first is located around 47 meV along the $R$–$A$–$Z$ directions and originates from the vibrations of O atoms in the $xy$-plane. The second corresponds to the highest optical mode along the $\Gamma$–$Z$ direction, arising from the half-breathing mode vibrations of O atoms coupled with the vibrations of Ni atoms in the $xy$-plane. Flatness along $\Gamma-Z$ suggests that these phonons are largely two-dimensional (2D), which could enhance the EPC in LaNiO$_2$ because 2D phonons couple more effectively with electrons in layered materials~\cite{Giant_ephSrRuO3}.

The Eliashberg spectral function $\alpha^{2}F(\omega)$ and the cumulative EPC strength $\lambda$ in the NM phase, are depicted in Fig.~\ref{fig:nm_epc}(c). It can be easily seen that the EPC strength is contributed by phonon modes from all atomic vibrations, not solely from the full- and half-breathing modes of O. The two dominant broad features contributing to $\alpha^{2}F(\omega)$ are located around $\sim$60 and 35 meV. The former primarily originates from the vibrations of O atoms in the $xy$-plane, including the half-breathing mode at the X point [labeled as X$_1$ mode in Fig.~\ref{fig:nm_epc}(d)], and the anti-phase half-breathing mode at the R point [labeled as R$_1$ mode in Fig.~\ref{fig:nm_epc}(d)], and several phonon modes along the $M–\Gamma$ and $R–A$ lines, which result from the breathing O modes coupled with the vibrations of Ni atoms in the $xy$-plane. The second broad peak is dominated by the breathing modes of La atoms along the $z$-direction at the Z point [labeled as $Z_1$ mode in Fig.~\ref{fig:nm_epc}(d)], which produces the second most prominent peak in $\alpha^{2}F(\omega)$. Furthermore, the shear mode of Ni atoms at the X point [labeled as $X_2$ mode in Fig.~\ref{fig:nm_epc}(d)] contributes significantly, forming the third prominent peak in $\alpha^{2}F(\omega)$ near 20 meV. The cumulative EPC strength \(\lambda\) is found to be 0.16, resulting in \(T_c\) being zero with \(\mu^{*}=0.1\) based on Eq.~\ref{eq:tc}. These results indicate that the EPC of the NM phase is very weak, consistent with previous DFT studies~\cite{Nomura_eph_LNO}. 

To understand the origin of EPC in the NM phase, we present the FS nesting function at $q_z = 0$ (left panel) and $q_z = \pi/c$ (right panel) in Fig.~\ref{fig:nm_epc}(e). According to Eq.~\ref{eq:phse}, the phonon self-energy depends on two key terms: the EPC $g$-matrix elements and the bare electronic susceptibility, $\chi_0(\mathbf{q},\omega)$. The FS nesting function is simply the static limit of the imaginary part of the susceptibility, which reflects purely electronic contributions. Several pronounced nesting peaks are clearly visible in both the $k_z = 0$ and $k_z = \pi/c$ planes [Fig.~\ref{fig:nm_epc}(e)]. Specifically, strong nesting features appear near the wave vectors $q=(0.25,0,0)$, $q=(0.25,0.25,0)$, and $q= (0.5,0,0)$, indicating that electronic states near these vectors are sensitive to charge density modulation.

The $k$-resolved EPC strength, $\lambda_{k}$, is plotted in Fig.~\ref{fig:nm_epc}(f) on the $k_z = 0$ (left panel) and $k_z = \pi/c$ (right panel) planes. Comparing these plots with the FS projections in Fig.~\ref{fig:elec}(c) reveals that all orbitals crossing the Fermi level contribute significantly to the EPC in the NM phase. On the $k_z = \pi/c$ plane, broadened peaks centered around the R point correspond to the VHS in the band structure [Fig.~\ref{fig:elec}(a)]. This suggests that the VHS plays a critical role in enhancing the EPC in this $k_z$ plane. Additionally, the large electron pocket centered at the R point, primarily contributed by the Ni $3d_{xz/yz}$ orbitals, shows stronger EPC compared to other orbitals, indicating that orbital contributions are highly anisotropic and dependent on FS topology. 

\SB{Evaluation of EPC for the $C$-AFM phase}\textbf{\---}We now turn to the discussion of EPC in the $C$-AFM phase of pristine LaNiO$_2$. The folded phonon dispersion weighted by the phonon-mode-resolved EPC $\lambda{\textbf{q}\nu}$ along the high-symmetry lines of the NM BZ, along with the phonon density of states (PhDOS) and the Eliashberg spectral function $\alpha^{2}F(\omega)$, are shown in Fig.~\ref{fig:afm_epc}(a)–(c). Unlike the NM case, the $\lambda_{\textbf{q}\nu}$ values in the $C$-AFM phase are significantly enhanced in the $k_z = 0$ plane but vanish in the $k_z = \pi/c$ plane. This can be attributed to the confinement of the FS in the $k_z = \pi/c$ plane, which results in strong FS nesting in the $q = 0$ plane, while leaving no nesting in the $q_z = \pi/c$ plane [Fig.~\ref{fig:afm_epc}(e)]. Therefore, the phonon softening of $C$-AFM is due to the synergistic effects from the strong EPC and FS nesting.

Figure~\ref{fig:afm_epc}(c) compares the isotropic Eliashberg spectral function $\alpha^{2}F(\omega)$ of the $C$-AFM (red) and NM (yellow) phases.  Both the high and the low frequency modes are seen to be enhanced in the $C$-AFM phase, compared to the NM case. In particular, the peaks in the low-frequency region near $\sim$15 meV are greatly enhanced.  The two key phonon modes involved here are as follows. (i) The breathing mode of La atoms along the [110] direction [$X_2$ in Fig.~\ref{fig:afm_epc}(d)], with a frequency of 17.0 meV, indicating the critical role of rare earth atoms in the IL-nickelates. And (ii) the 12.9 meV mode, which results from the shear vibrations of Ni atoms along the [001] direction [$X_3$ in Fig.~\ref{fig:afm_epc}(d)] and contributes significantly to the total EPC. In this mode, the nearest-neighbor spin-up Ni atoms exhibit shear displacements, while the spin-down Ni atoms remain stationary. These results show that low-frequency lattice vibrations, particularly those involving La and Ni atoms, are essential for understanding the EPC in the $C$-AFM phase. 

The cumulative EPC strength $\lambda$ is found to be 0.66, which is 4 times larger than the NM value of 0.16. According to Eqs.~\eqref{eq:tc} and \eqref{wlog}, the estimated $\omega_{log}$ is $\sim$17.9 meV, and the predicted $T_\text{c}$ ranges from 5.1$-$7.3 K, with the adjustable parameter $\mu^{*}$ set between 0.12 and 0.08, respectively. For 5\% hole doping, using the rigid-band approximation, we estimate that $\lambda$ will be further enhanced to 0.89, leading to $T_\text{c}$ ranges from 10.7$-$13.4 K with $\mu^{*}$ set between 0.12 and 0.08, respectively [see the section 2 of Supplemental Material (SM) for details]. It should be noted that for such large values of $\lambda$ the superconducting gap equation should be solved to obtain a more accurate value of the $T_c$~\cite{Anisotropic_ME}.

Within the double-delta approximation, the phonon self-energy in Eq.~\ref{eq:phse_doubledelta} receives contributions only from the electronic states near the Fermi level, which originate from the two very flat pockets near the $k_z = \pi/c$ plane [Fig.~\ref{fig:elec} (e)]. One is an electron-like pocket centered at the $Z$ point, which arises from strong hybridization between the Ni $d_{z^2}$ and $d_{xz/yz}$ orbitals. The other is a hole-like pocket centered at the $R$ point, mainly contributed by Ni $d_{z^2}$ orbitals. To identify which orbitals are crucial for the EPC in the $C$-AFM phase, we plot the $k$-resolved EPC strength $\lambda_{k}$ for the $k_z=0$ (left panel) and $k_z = \pi/c$ planes (right panel) in Fig.~\ref{fig:afm_epc}(f). Notably, the $k_z = \pi/c$ plane in which the flat bands are located shows a significant enhancement in $\lambda_{k}$. This highlights that the Ni $d_{z^2}$ and $d_{xz/yz}$ orbitals, which dominate the flat bands, play a central role in driving strong EPC in the $C$-AFM phase. In contrast, the $k_z = 0$ plane shows no EPC contribution, which is consistent with the absence of FS nesting at $k_z = 0$ [Fig.~\ref{fig:elec} (b)]. These results indicate that strong EPC is closely correlated with the FS topology, particularly the pronounced nesting and flat bands at $k_z = \pi/c$ [Fig.~\ref{fig:afm_epc} (e)].

\begin{figure}[t!]
\centering
\includegraphics[width=.98\columnwidth]{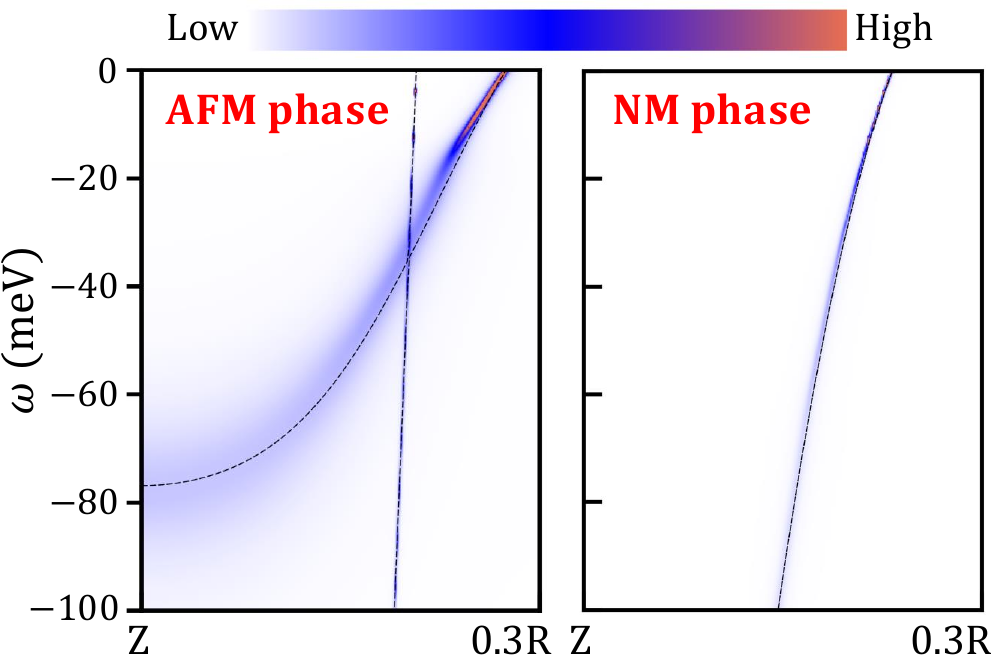}
\caption{Calculated EPC-renormalized electron spectral functions, $A_{n\boldsymbol{k}}(\omega)$, for the $C$-AFM (a) and NM (b) phases along the $Z$–$R$ direction. The dashed black lines represent the bare electronic bands obtained from DFT calculations.}
\label{fig:kink_cmp}
\end{figure}

\SB{Kinks in the IL-nikcleats}\textbf{\---}To gain more insights into the renormalization of the electronic structure of IL-nickelate due to the EPC, we calculated the electron spectral functions $A_{nk}(\omega)$ of IL-nickelate based on Eqs.~\ref{FM_self} and~\ref{Akw}. Figure~\ref{fig:kink_cmp} compares the calculated spectral functions $A_{nk}(\omega)$ of the $C$-AFM (left panel) and NM (right panel) phases along the $Z$–$R$ direction. Remarkably, a distinct kink emerges in the $C$-AFM phase near a binding energy of approximately 15 meV, which aligns closely with the energy of prominent low-frequency phonon modes identified above. These quasipaticle self-energy features constitute clear hallmarks of strong EPC present in the $C$-AFM phase. In addition, we see a large broadening corresponding to strong EPC with high-energy phonon modes, although no kink is observed. In contrast, no kink or significant broadening is observed in the NM phase, consistent with its weak EPC discussed above.  The flat bands and magnetism thus play a critical role for understanding the EPC in LaNiO$_2$.

\SB{Discussion}\textbf{\---}Many experiments show that IL-nickelates share several key characteristics with the cuprates~\cite{Lee2023_linear_tc,SC_dome_NNO,Phase_D_NdNiO2,Lin_2022,OrtizPRR2022,Lu2021science,ZhangPRLLaNiO2}, suggesting that their $T_c$ is driven by strong electron correlations. The pronounced phonon softening and strong EPC predicted here in the magnetic phase of nickelates indicate that phonons may also play a significant role. The recently observed low-energy kinks around 70-100 meV in the hole-doped LaNiO$_2$ from ARPES experiments are likely a signature of strong EPC~\cite{Sun_ARPES_LNO} as electronic correlation effects alone cannot explain this phenomenon~\cite{Hardening_YWang,Sun_ARPES_LNO,Liang_DMFT_LNO}. Although the observed FS from ARPES ~\cite{Sun_ARPES_LNO} is in accord with the DFT-based predictions for the NM phase, the small value of EPC in the NM phase is insufficient to account for the kink observed in the hole-doped LaNiO$_2$. The DFT also predicts the NM phase to be a good metal, which contradicts the weakly insulating behavior observed in underdoped and pristine LaNiO$_2$. In contrast, the scenario of competing phases in pristine LaNiO$_2$~\cite{ZhangPRLLaNiO2} provides a more plausible explanation for the weakly insulating state. Our finding that the EPC is enhanced substantially in the $C$-AFM phase and  that it induces a prominent kink in the electronic spectrum around 15 meV, indicates the EPC plays a key role in the physics of the hole-doped nickelates. Interestingly, the presence of low- and high-energy kinks, along with vestiges of non-magnetic FS are also found in the cuprates~\cite{Seppo,Lanzara,LEK}. Note that, although our current calculations focus primarily on pristine LaNiO$_2$,  flat bands in the low-energy stripe phases are distributed over a wide range of energies due to the symmetry-lowering, as demonstrated in our previous study~\cite{ZhangPRLLaNiO2}. This strongly suggests that flat bands should persist near the Fermi level upon hole doping. In this connection, we expect that strong EPC could also manifest prominently in the hole-doped nickelates. However, a more detailed investigation into the EPC of doped LaNiO$_2$ remains necessary and will be addressed in our future studies. 

One of our key findings is that the magnetic order can produce enhancement of EPC of a wide variety of phonon modes.  In LaNiO$_2$, we show that this enhancement is driven mainly by the low-energy phonon modes which involve vibrations of La and Ni atoms [Fig.~\ref{fig:afm_epc}]. We also find that the full-breathing O modes actually harden in $C$-AFM, although these bond disproportionating vibrations have been implicated in this connection in previous studies~\cite{CO_elph_Julien,COBO_elph_Julien}.  Note here that  a high energy optical phonon cannot simply soften to zero frequency via a Kohn anomaly, but in doing so it must anticross other lower energy modes having the same symmetry. Anticrossing effects, which segment the Kohn anomaly into a ladder of softened phonon branches have been discussed in connection with monolayer NbSe$_2$ and the cuprates~\cite{NbSe2_Susy,Markiewicz_2015,Zhou_phonon,Huang_fluctuations_cuprates,LEK}.  This anticrossing effects the transfer of breathing mode phonons to lower phonon branches, thereby reconciling their strong role in enhancing EPC with the observed hardening of the original breathing-mode branch.

\SB{Conclusion}\textbf{\---}Our in-depth, material-specific modeling of EPC effects in pristine LaNiO$_2$, where the spin, orbital, charge, and lattice degrees of freedom are treated on an equal footing without invoking any free parameters such as the Hubbard $U$, reveal that EPC is significantly enhanced by the presence of magnetism. This enhancement is driven by the flat bands in the undoped nickelate and results in a dramatic softening of the phonons and is predicted to induce a signature kink in the renormalized electronic spectrum of the $C$-AFM phase at around 15 meV. Our study suggests the value of further theoretical and experimental work towards unraveling the effects of EPC in the infinite-layer nickelates.

\SB{Methods}\textbf{\---}First-principles calculations were performed using the projector-augmented-wave (PAW) method~\cite{Kresse1999}, as implemented in the Vienna Ab initio Simulation Package (VASP)~\cite{Kresse1993,Kresse1996}. A high-energy cutoff of 600 eV was applied to truncate the plane-wave basis set. Crystal structures and ionic positions were fully optimized with a force convergence criterion of 0.001 eV/\AA{} for each atom and a total energy tolerance of $10^{-8}$ eV. For the NM phase of LaNiO$_2$, a $\Gamma$-centered k-point mesh of $10 \times 10 \times 12$ was employed for structural relaxation. Phonon calculations were carried out using a $6 \times 6 \times 6$ supercell with a $\Gamma$-centered k-point mesh of $3 \times 3 \times 3$, employing the finite displacement method. A Gaussian smearing with a small $\sigma$ of 0.02 eV was applied in the phonon calculations. To simulate the $C$-AFM phase, a $\sqrt{2} \times \sqrt{2} \times 1$ supercell was used. Structural relaxation of the $C$-AFM phase was performed using a $\Gamma$-centered k-point mesh of $9 \times 9 \times 15$. Phonon calculations for the $C$-AFM phase were performed using a $4 \times 4 \times 4$ supercell of $C$-AFM unit cell with a $\Gamma$-centered $k$-point mesh of $2 \times 2 \times 3$, using the finite displacement method. The Wannier representation of the electronic band structure was obtained using Wannier90~\cite{MOSTOFI20142309} with the help of the VASP2WANNIER90 interface. La $f$ and $d$, and Ni $d$ states were included for constructing the Wannier model [see section 1 of SM]. Electron-phonon coupling matrix elements were computed using the same computational strategy outlined in Ref.~\cite{Manuel2020_epc}, but with the all-electron matrix elements as defined in Ref.~\cite{FD_PAW}.

Previous SCAN/\rrscan studies have demonstrated their efficacy (compared to PBE and LDA) in providing an accurate description of total energies and electronic structures for complex materials containing $d$- and $f$-electrons~\cite{Zhang2020b,ZhangR2020}, including the cuprates~\cite{Lane2018, Furness2018, ZhangYPANS} and nickelates~\cite{Zhang2021_nickelate,Lane2023,ZhangPRLLaNiO2}. Recent work has also shown that SCAN/\rrscan can accurately capture phonon dispersions in the cuprates~\cite{Zhang2020b, NingPRBYBCO6} and NiO~\cite{NingCMPhonon}. More recently, we have shown that \rrscan provides an improved description of EPC in challenging transition-metal oxides like CoO and NiO~\cite{eph_r2scan_MO}. Given that \rrscan exhibits better numerical stability than SCAN~\cite{R2SCANJPCL}, we employed the \rrscan functional for all calculations in this study.

EPC strength $\lambda$ and the isotropic Eliashberg spectral function $\alpha^{2}F$ were computed as follows. Within the Migdal approximation~\cite{Migdal1958}, the imaginary part of the phonon self-energy is: 

\begin{equation}\label{eq:phse}
\Pi''_{\vecq\nu} = \text{Im} \sum_{m,n,\veck} w_{\veck}  \vert g_{mn,\nu}( {\veck,\vecq} ) \vert^2 
\frac{ f_{n\veck} - f_{m{\veck+\vecq}} }{
\epsilon_{m{\veck+\vecq}} - \epsilon_{n\veck} - \omega_{\vecq\nu} - i\eta},
\end{equation}
where $f_{n{\veck}}$ is the Fermi occupation of the single-particle state $\vert\psi_{n\veck}\rangle$ associated with the eigenvalue $\epsilon_{n\veck}$, $w_{\veck}$ are the weights of the dense ${\veck}$-points, and $\eta$ indicates a real positive infinitesimal. We note here that the spin index is omitted for simplicity. The EPC strength $\lambda$ is the full BZ average of the mode-resolved $\lambda_{\vecq\nu}$
\begin{equation}\label{eq:lam}
\lambda = \sum_{\vecq\nu} w_{\vecq} \lambda_{\vecq\nu},
\end{equation}
with $w_{\vecq}$ being the weights of the dense $\vecq$-points. We adopted the double-delta approximation~\cite{epiq_impl2024} to calculate the imaginary part of the phonon self-energy, under which Eq.~\eqref{eq:phse} can be rewritten as:
\begin{equation}\label{eq:phse_doubledelta}
\Pi''_{\vecq\nu} = \sum_{m,n,\veck} w_{\veck} \omega_{\vecq\nu} \vert g_{mn,\nu}( {\veck,\vecq} ) \vert^2 \delta(\epsilon_{n{\bf k}} -\epsilon_F ) \delta(\epsilon_{m\veck+\vecq} - \epsilon_F ),
\end{equation}
and $\lambda_{\vecq\nu}$ can be expressed as:
\begin{equation}
\lambda_{\vecq\nu} = \frac{1}{\pi N_\text{F}} \frac{ \Pi''_{\vecq\nu}}{ \omega^2_{\vecq\nu}},
\end{equation}
where $N_\text{F}$ is the density of states at the Fermi level. Using this mode-resolved $\lambda_{\vecq\nu}$ and the phonon frequencies $\omega_{\vecq\nu}$, the isotropic Eliashberg spectral function, $\alpha^2 F$, can be calculated by using
\begin{equation}\label{eq:a2F}
\alpha^2F(\omega) = \frac{1}{2}\sum_{\vecq\nu}  
w_{\vecq} \omega_{\vecq\nu} \lambda_{\vecq\nu} \, \delta( \omega - \omega_{\vecq\nu}).
\end{equation}
The phonon-mediated superconducting transition temperature, $T_\text{c}$, is evaluated with the McMillian-Allen-Dynes formula~\cite{McMillan1968},
\begin{equation}
T_\text{c} = \frac{\omega_\text{log}}{1.2}\exp\left[{\frac{-1.04(1+\lambda)}{\lambda(1 - 0.62\mu^*) - \mu^*}}\right],
\label{eq:tc}
\end{equation}
where the logarithmic average frequency $\omega_{\text{log}}$ can be calculated from the following equation~\cite{Allen1975,grimvall1981electron}
\begin{equation}\label{wlog}
\omega_\text{log} = \exp\left[\frac{2}{\lambda}\int^\infty_0\text{d}\omega \frac{ \log\,\omega}{\omega}\alpha^2F(\omega)\right].
\end{equation}

The fermi-nesting function, which describes electron-hole excitations around the fermi surface, is evaluated by using
\begin{align}
\chi_\text{nest}(\boldsymbol{q}) &= \sum_{m,n,\boldsymbol{k}} \delta(\epsilon_{n\boldsymbol{k}} - \epsilon_F) \delta( \epsilon_{m{\boldsymbol{k+q}}} -\epsilon_F ).
\end{align}

To obtain the electronic spectral function, we also calculated the Fan-Migdal term, which reflects the electron-phonon interaction contributions to the electronic self-energy, using the equation:
\begin{align}
&\Sigma^\text{FM}_{n\veck}(\omega,\, T) = \sum_m\sum_{\vecq\nu} w_{\vecq} \left\vert g_{mn,\nu}(\veck,\,\vecq)\right\vert^2 \times \notag \\
&\quad\left[ \frac{ 1- f_{m\veck+\vecq}(T) + n_{\vecq\nu}(T) }{ \omega -\epsilon_{m\veck+\vecq} -\omega_{\vecq\nu} + i\eta} + \frac{ f_{m\veck+\vecq}(T) + n_{\vecq\nu}(T) }{ \omega -\epsilon_{m\veck+\vecq} +\omega_{\vecq\nu} + i\eta} \right],
\label{FM_self}
\end{align}
where $f_{m\veck+\vecq}(T)$ and $n_{\vecq\nu}(T)$ are the Fermi-Dirac and Bose-Einstein distribution functions for the electrons and phonons of the system at temperature $T$, respectively. The infinitesimal positive quantity $\eta$ is introduced to ensure physical causality and as an empirical tuning parameter in the calculations~\cite{Gonze_Marini2015/JCP}. The electronic spectral function is then obtained from the calculated phonon-induced self-energy using the following equation:
\begin{align}
A_{\veck}(\omega) &= -\frac{1}{\pi} \sum_{n} \cfrac{ \text{Im}\Sigma^\text{e-ph}_{n\veck}(\omega) }{ \Big[ \omega - \epsilon_{n\veck} -\text{Re}\Sigma^\text{e-ph}_{n\veck}(\omega) \Big]^2 +\left[ \text{Im}\Sigma^\text{e-ph}_{n\veck}(\omega) \right]^2}.
\label{Akw}
\end{align}

Here, we subtract $\text{Re}\Sigma^\text{e-ph}_{n\veck}(E_F)$ from the calculated $\text{Re}\Sigma^\text{e-ph}_{n\veck}(\omega)$ to ensure electron number conservation~\cite{FelicianoRMP2017,epwcpc} according to Luttinger's theorem~\cite{Luttinger_1960}.

~\\

\bibliography{Ref_nourl}

\SB{Acknowledgements}\textbf{\---}R.Z., Y.W., and J.S. acknowledge support from the U.S. Office of Naval Research (ONR) Grant No. N00014-22-1-2673.  The work at Tulane University was also supported by the Advanced Cyberinfrastructure Coordination Ecosystem funded by the National Science Foundation, and the National Energy Research Scientific Computing Center (NERSC) using NERSC Awards Nos. BES-ERCAP0032902 and BES-ERCAP0032886. The work at Northeastern University was supported by the National Science Foundation through the Expand-QISE award NSF-OMA-2329067 and benefited from the resources of Northeastern University’s Advanced Scientific Computation Center, the Discovery Cluster, the Massachusetts Technology Collaborative award number MTC-22032, and the Quantum Materials and Sensing Institute. The work at Los Alamos National Laboratory was carried out under the auspices of the U.S. Department of Energy (DOE) National Nuclear Security Administration under Contract No. 89233218CNA000001. It was supported by the LANL LDRD Program, the Quantum Science Center, a U.S. DOE Office of Science National Quantum Information Science Research Center, and in part by the Center for Integrated Nanotechnologies, a DOE BES user facility, in partnership with the LANL Institutional Computing Program for computational resources. Additional computations were performed at the NERSC, a U.S. Department of Energy Office of Science User Facility located at Lawrence Berkeley National Laboratory, operated under Contract No. DE-AC02-05CH11231 using NERSC Award No. ERCAP0020494. The work at Howard University was supported by the National Science Foundation through the Expand-QISE award NSF-OMA-2329067 and the resources of Accelerate ACCESS PHYS220127 and PHYS2100073. B. B. acknowledges support from the COST Action
CA16218.

\SB{Author contributions}\textbf{\---}R.Z., J.S., and A.B. designed the study. R.Z., Y.W, and M.E. performed first principles computations and analyzed the data with help from C.L. H.M., L.H., S.C, B.S, B.B, J.Z., R.S.M., E.K.U.G., G. K., A.B., and J.S., A.B., and J.S. provided computational infrastructure. R.Z., Y.W, C.L., B.S., B.B, R.S.M., A.B. and J.S. wrote the manuscript with input from all the authors. All authors contributed to editing the manuscript.

\SB{Additional information}\textbf{\---}The authors declare no competing financial interests. 

\end{document}